\documentclass[aps,superscriptaddress,showpacs,preprint,amsmath,amssymb]{revtex4}
\usepackage{graphicx}
\usepackage{epsfig}
\usepackage{dcolumn}
\usepackage{bm}
\usepackage{subfigure}

\newcommand{\chisq}[1]{$\chi^{2}_{#1}$}

\def \jp {J/\psi}

\def \ee {e^+e^-}
\def \kk {K^+K^-}

\def \rhopi {\rho\pi}
\def \kkp {\kk\pi^0}
\def \gg {\gamma\gamma}
\def \chisq {\chi^{2}}
\newcommand{\chicj}{\chi_{cJ}}

\newcommand{\psip}{\psi^{\prime}}

\begin{document}
\title{\quad\\[1.0cm] \boldmath
Experimental study of $\psip$ decays to $\kk\pi^0$ and $\kk\eta$}

\author{
\small{
M.~Ablikim$^{1}$, M.~N.~Achasov$^{5}$, O.~Albayrak$^{3}$, D.~J.~Ambrose$^{39}$, F.~F.~An$^{1}$, Q.~An$^{40}$, J.~Z.~Bai$^{1}$, Y.~Ban$^{27}$, J.~Becker$^{2}$, J.~V.~Bennett$^{17}$
% ...
\\
\vspace{0.2cm}
(BESIII Collaboration)\\
\vspace{0.2cm} {\it
$^{1}$ Institute of High Energy Physics, Beijing 100049, P. R. China\\
$^{2}$ Bochum Ruhr-University, 44780 Bochum, Germany\\
$^{3}$ Carnegie Mellon University, Pittsburgh, PA 15213, USA\\
% ...
$^{5}$ G.I. Budker Institute of Nuclear Physics SB RAS (BINP), Novosibirsk 630090, Russia\\
% ...
$^{17}$ Indiana University, Bloomington, Indiana 47405, USA\\
% ...
$^{27}$ Peking University, Beijing 100871, P. R. China\\
% ...
$^{39}$ University of Rochester, Rochester, New York 14627, USA\\
$^{40}$ University of Science and Technology of China, Hefei 230026, P. R. China\\
% ...
}
}
}
\begin{abstract}
Using $(106\pm4)\times 10^6$ $\psip$ events accumulated with the
BESIII detector at the BEPCII $e^+e^-$ collider, we present
measurements of the branching fractions for $\psip$ decays to
$~\kk\pi^0$ and $\kk\eta$. In these final states, the decay $\psip\to
K_2^{*}(1430)^+K^-+c.c.$ is observed for the first time, and its
branching fraction is measured to be $(7.12\pm{0.62}{\rm
~(stat.)}^{+1.13}_{-0.61}{\rm~(syst.)})\times 10^{-5}$, which indicates a
violation of the helicity selection rule in $\psip$ decays. The
branching fractions of $\psip\to
K^*(892)^+K^-+c.c.,~\phi\eta,~\phi\pi^0$ are also measured.  The
measurements are used to test the QCD predictions on charmonium
decays.
\end{abstract}
\pacs{13.25.Gv, 13.20.Gd, 14.40.Pq}
\maketitle
\section{Introduction}
 In the framework of perturbative QCD (pQCD), $\jp$ and $\psip$ decays
 to light hadrons are expected to be dominated by the annihilation of
 $c\bar c$ quarks into three gluons or one virtual photon, with hadron
 decay partial widths that are proportional to the square of the
 $c\bar c$ wave function overlaps at the origin, which can be related
 to their leptonic decay widths \cite{tpr}. This suggests that the
 ratio $Q_h$ of branching fractions for $\psip$ and $\jp$ decays to
 the same final state should follow the rule:
\begin{eqnarray}\label{qh_definition}
  Q_{h}=\frac{Br(\psip\rightarrow h)}{Br(J/\psi\rightarrow h)}\cong \frac{Br(\psip\rightarrow e^+e^-)}{Br(J/\psi\rightarrow e^+e^-)}\cong 12\%,
\end{eqnarray}
where $Br$ denotes a branching fraction and $h$ is a particular
hadronic final state. This relation is referred to as the ``12\%
rule".

Although the 12\% rule works well for some specific decay modes of the $\psip$, the decay $\psip$ to $\rho\pi$ exhibits a factor of 70 times stronger suppression than expectations based on this rule. This suppression in vector-pseudoscalar (VP) meson modes was first observed by MARKII \cite{kkpi0},  which is referred to as the ``$\rhopi$ puzzle". Further tests of this rule in the VP modes have been performed by CLEO \cite{cleoVP} and BESII \cite{bes2VP}, and have been extended to the pseudoscalar-pseudoscalar meson (PP), vector-tensor meson (VT) and multibody decays.  Although $Q_h$ values have been measured for a wide variety of final states, most of them have large uncertainties due to low statistics \cite{pdg}. Reviews of the rho-pi puzzle conclude that
current theoretical explanations are unsatisfactory \cite{guyf}. More experimental results are desirable.

For charmonium $\psi(\lambda)$ decays to light hadrons $h_1(\lambda_1)$ and $h_2(\lambda_2)$, the asymptotic behavior of the branching fraction from a pQCD calculation to leading twist accuracy gives \cite{charmoniumBr}:
 \begin{equation}
Br[\psi(\lambda)\to h_1(\lambda_1)h_2(\lambda_2)]\sim \left({\Lambda^2_\textrm{QCD} \over m_c^2} \right)^{|\lambda_1+\lambda_2|+2},
 \end{equation}
 where $\lambda,~\lambda_1$ and $\lambda_2$ denote the helicities of the corresponding hadrons. Here $m_c$ is the charm quark mass and $\Lambda_\textrm{QCD}$ is the QCD energy scale factor. If the light quark masses are neglected, the vector-gluon coupling conserves quark helicity and this leads to the helicity selection rule (HSR) \cite{hsr}: $\lambda_1+\lambda_2=0$. If the helicity configurations do not satisfy this relation, the branching fraction should be suppressed.

For the $\psip$ decays to VP [$K^*(892)^\pm K^\mp$] or TP [$K^*_2(1430)^\pm K^\mp$], the amplitudes are antisymmetric in terms of the final state helicities, since strong or electromagnetic interactions conserve parity. Hence the amplitudes vanish when $\lambda_1=\lambda_2=0$. Nonvanishing amplitudes require the helicity configuration to satisfy the relation $|\lambda_1+\lambda_2|=1$, which violates the HSR and the branching fractions are expected to be suppressed.

Strikingly, HSR-violating decays were recently observed in $\chicj$ decays into vector-vector meson pairs  by BESIII \cite{chicj2vv}, which strongly indicates the failure of the HSR \cite{zhoaqiang}. In an analysis of $\psip\to K_S^0K^\pm\pi^\mp$ by BESII~\cite{bes2VP}, evidence for $\psip\to K^{*}_J K^0$ ($K^{*}_J$ refers to either $K^{*}_J(1430)$ or $K^{*}(1410)$) was seen, but low statistics prevented a further study.

With the large $\psip$ data sample accumulated by the BESIII
experiment, new opportunities to precisely test the $12\%$ rule in the
decays of $\psip\to K^{*}(892)^+ K^-+c.c.$ and $\eta\phi$, and to
search for $\psip\to K_2^{*}(1430)^\pm K^{\mp}$ are available. Such
measurements can shed light on charmonium decay mechanisms and,
therefore, be helpful for understanding the $\rho\pi$ puzzle. In
particular, the decay $\psip\to\kk\eta$ provides opportunities to
study not only $\phi \eta$, but also the excited $\phi$
states, such as $\phi_3(1850)$ and $\phi(2170)$. The decay $\psip\to
\kkp$ also allows us to study the isospin violation decay
$\psip\to\phi\pi^0$, which is expected to proceed via electromagnetic
(EM) processes \cite{Wang:2012mf}.

\section{The BESIII experiment and data set}
We use a data sample containing $(106\pm4)\times 10^6$ $\psip$ decays
recorded with the BESIII detector \cite{bes3detector} at the energy-symmetric
double ring $\ee$ collider BEPCII. The primary
data sample corresponds to an integrated luminosity of
156.4 pb$^{-1}$ collected at the peak of the $\psip$ resonance. In
addition, a 2.9 fb$^{-1}$(43 pb$^{-1}$) data sample collected at a center-of-mass energy of 3.773 GeV (3.65 GeV) is used for continuum background studies.

BEPCII is designed to
provide a peak luminosity of $10^{33}~$cm$^{-2}$s$^{-1}$ at a beam
current of 0.93 A for studies of hadron spectroscopy and $\tau-$charm physics \cite{yellowbook} . The BESIII detector is described in detail elsewhere \cite{bes3detector}.
Charged particle momenta are measured with a small-celled,
helium-gas-based main drift chamber (MDC) with 43 layers
operating within the 1T magnetic field of a solenoidal
superconducting magnet. Charged particle identification is
provided by measurements of the specific ionization energy
loss $dE/dx$ in the tracking device and by means of a
plastic scintillator time of flight (TOF) system comprised of a
barrel part and two endcaps. Photons are detected and
their energies and positions measured with an electromagnetic
calorimeter (EMC) consisting of 6240 CsI(Tl) crystals
arranged in a barrel and two endcaps. The return yoke
of the magnet is instrumented with resistive plate chambers
arranged in 9 (barrel) and 8 layers (endcaps) for the
discrimination of muons and charged hadrons.

The optimization of the event selection criteria and the estimation
of background sources are performed with
Monte Carlo~(MC)~simulated data samples. The {\sc geant4}-based simulation
software \cite{boost} includes the geometric and
material description of the BESIII detectors, the detector
response and digitization models, as well as the tracking of
the detector running conditions and performances. An inclusive $\psip$ MC sample is generated to study potential backgrounds. The
production of the $\psip$ resonance is simulated with
the MC event generator {\sc kkmc} \cite{kkmc}, while the
decays are generated with {\sc besevtgen} \cite{evtgen} for known decay
modes with branching fractions being set at their PDG \cite{pdg}
world average values, and with {\sc lundcharm} \cite{lundcharm} for the
remaining unknown decays. The analysis is performed in
the framework of the BESIII offline software system \cite{boss}
which provides the detector calibration, event reconstruction
and data storage.

\section{Event selection}
The selection criteria described below are similar to those used in previous BESIII analyses \cite{chicj2vv,chicj2gv}~and are optimized according to the signal significance.
\subsection{Photon identification}
Electromagnetic showers are reconstructed by clustering
EMC crystal energies. The energy deposited in nearby
TOF counters is included to improve the reconstruction
efficiency and the energy resolution. Shower identified as
photon candidates must satisfy fiducial and shower-quality
requirements. Photon candidates that are reconstructed
from the barrel region ($|\cos\theta| < 0.8$) must have a minimum
energy of 25 MeV, while those in the endcaps
($0.86 < | \cos\theta| < 0.92$) must have at least 50 MeV. Showers in the angular range between the barrel and endcap
are poorly reconstructed and excluded from the analysis.
To eliminate showers caused by bremsstrahlung charged particles, a photon
must be separated by at least $10^\circ$ from any charged track.
EMC cluster timing requirements are used to suppress
electronic noise and energy deposits from uncorrelated events.
The number of photon candidates $N_\gamma$ is required to be $2\le N_\gamma \le 10$.
\subsection{Charged particle identification}
Charged tracks are reconstructed from hits in the MDC. For each track, the
polar angle must satisfy $|\cos\theta| < 0.93$, and it must originate within
$\pm10$ cm from the interaction
point in the beam direction and within $\pm1$ cm of the
beam line in the plane perpendicular to the beam. The
number of charged tracks is required to be two with a net
charge of zero.
The time-of-flight and energy
loss $dE/dx$ measurements are combined to calculate particle
identification (PID) probabilities for pion, kaon, and
proton/antiproton hypotheses, and each track is assigned a
particle type corresponding to the hypothesis with the
highest confidence level. Both charged tracks are required
to be identified as kaons.

\subsection{Event selection criteria}

\begin{figure}
  \centering
  \includegraphics[width=10cm]{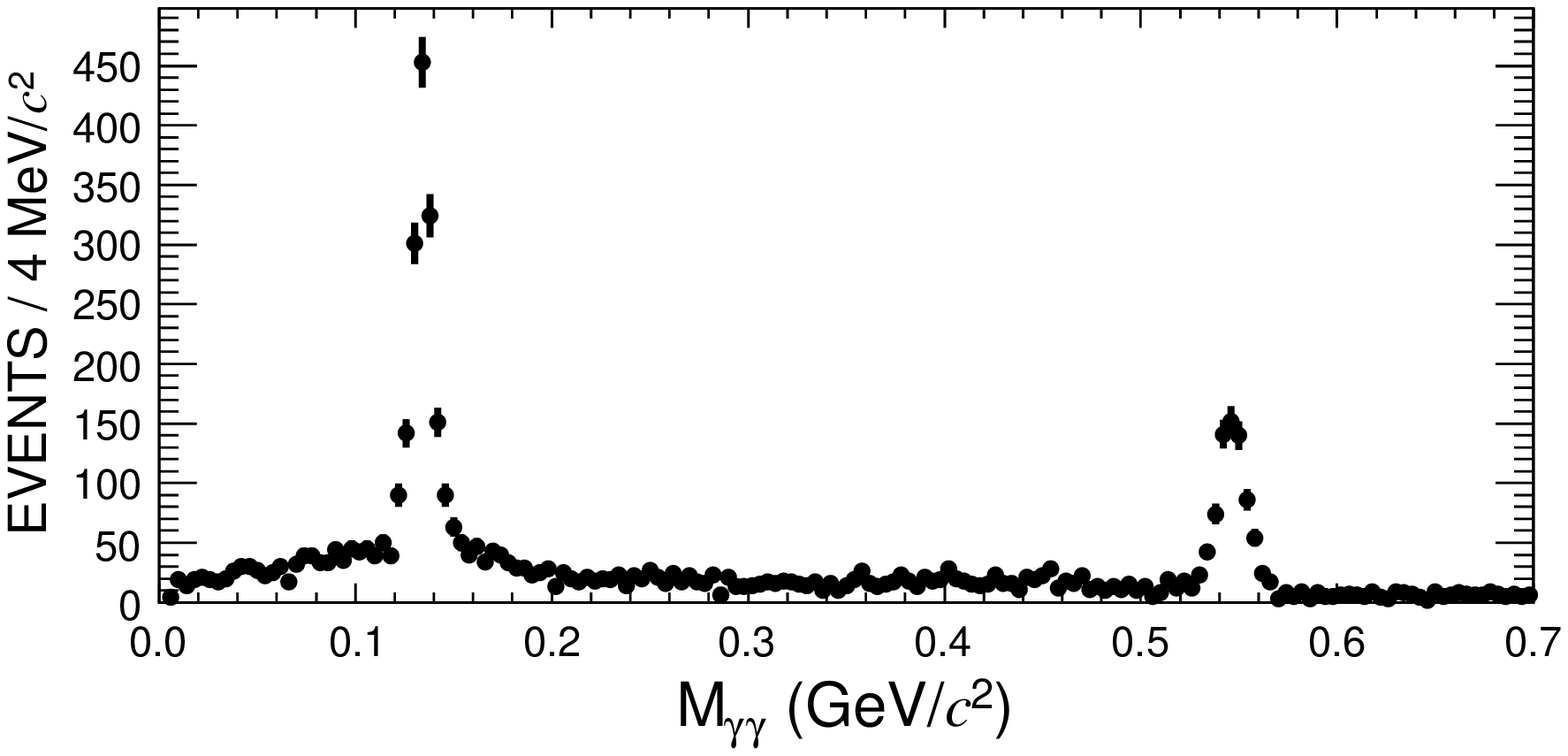}
  \caption{The invariant mass distribution for two photons in the selected
 $\psip\to\gg\kk$ events.}
  \label{mkk_vs_mgg}
\end{figure}

To choose the correct $\gamma\gamma$ combination for the $\pi^0$ or $\eta$
identification and to improve the overall mass resolution, a four-constraint kinematic
fit (4C-fit) is applied under the hypothesis $\psip\to\gamma\gamma \kk$ constrained
to the sum of the initial $\ee$ beam four-momentum. For events with more than two
photon candidates, the combination with the smallest $\chisq$
is kept. Candidates with $\chisq\le20$ for this fit are retained for further analysis.
Figure \ref{mkk_vs_mgg} shows the invariant mass distribution for the two selected
photons. Signal candidates of $\pi^0$ and $\eta$ mesons are clearly seen.

\subsubsection{\boldmath Final selection of $\psip\to\kk\pi^0$}

Candidates $\pi^0$ are selected by requiring the invariant mass of two photons, $M_{\gg}$, to
satisfy the condition $0.117 ~\textrm{GeV/c}^2\le M_{\gg}\le0.147 ~\textrm{GeV}/c^2$, an interval that is six times the $\pi^0$ mass resolution ($\sim$5 MeV$/c^2$). To suppress the background from
$\psip\to\gamma\chi_{c0}$, with $\chi_{c0}\to\kk$, it is required that the energy of the less
energetic photon ($E_{\gamma_{low}}$) is larger than 70 MeV. Background events from $\psip\to\pi^0\jp$,
with $\jp\to\kk$, are removed by requiring that the mass of the two kaons satisfies
$|M_{\kk}-m_{\jp}|\geq 7$~MeV/$c^2$, where $m_{\jp}$ is the $\jp$ mass \cite{pdg}.

\begin{figure}[hbtp]
  \centering
  \includegraphics[width=10cm]{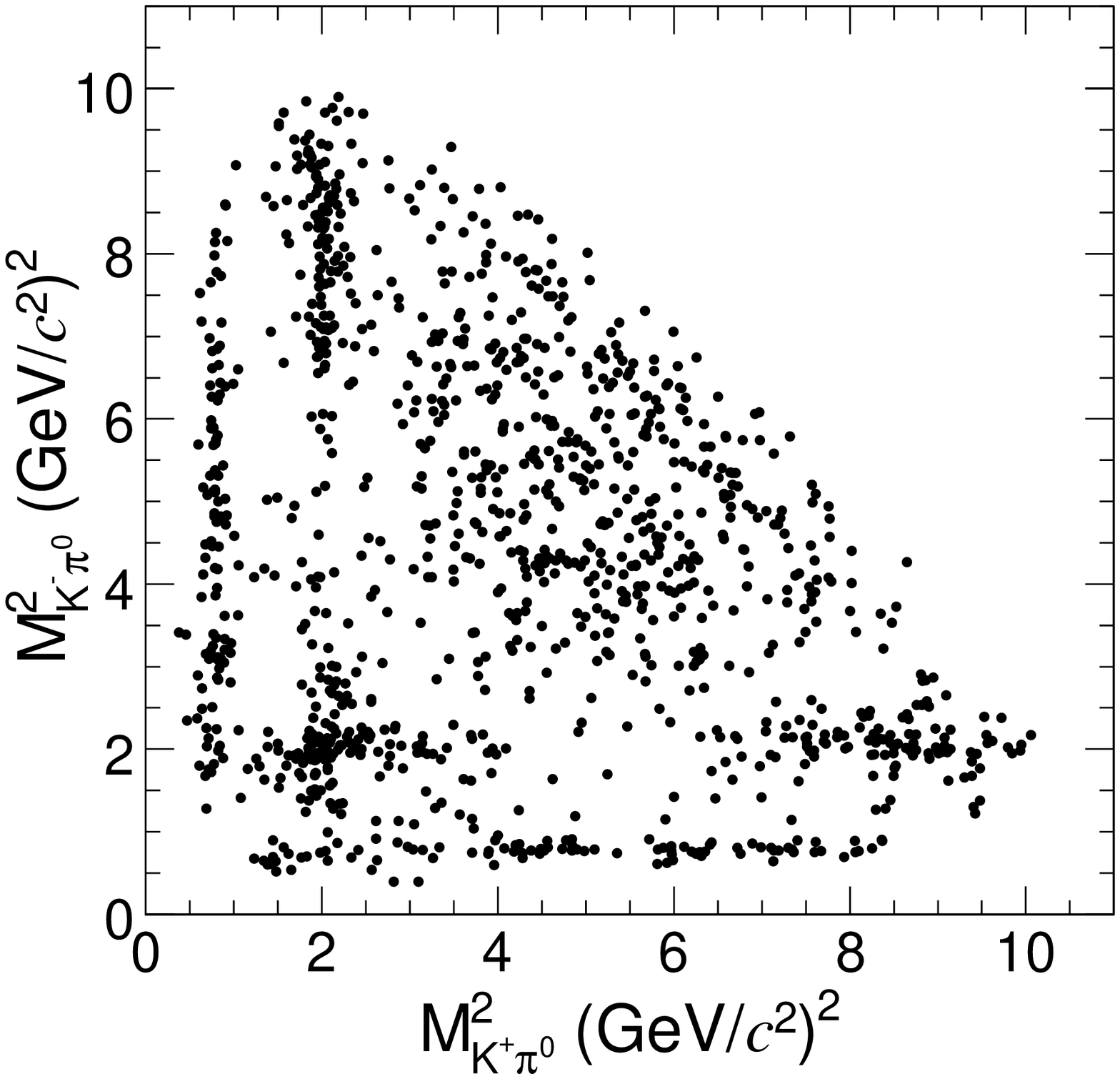}
  \caption{The Dalitz plot for $\psip\to\kkp$.}
  \label{dalitz}
\end{figure}

\begin{figure}[hbtp]
  \centering
  \includegraphics[width=10cm]{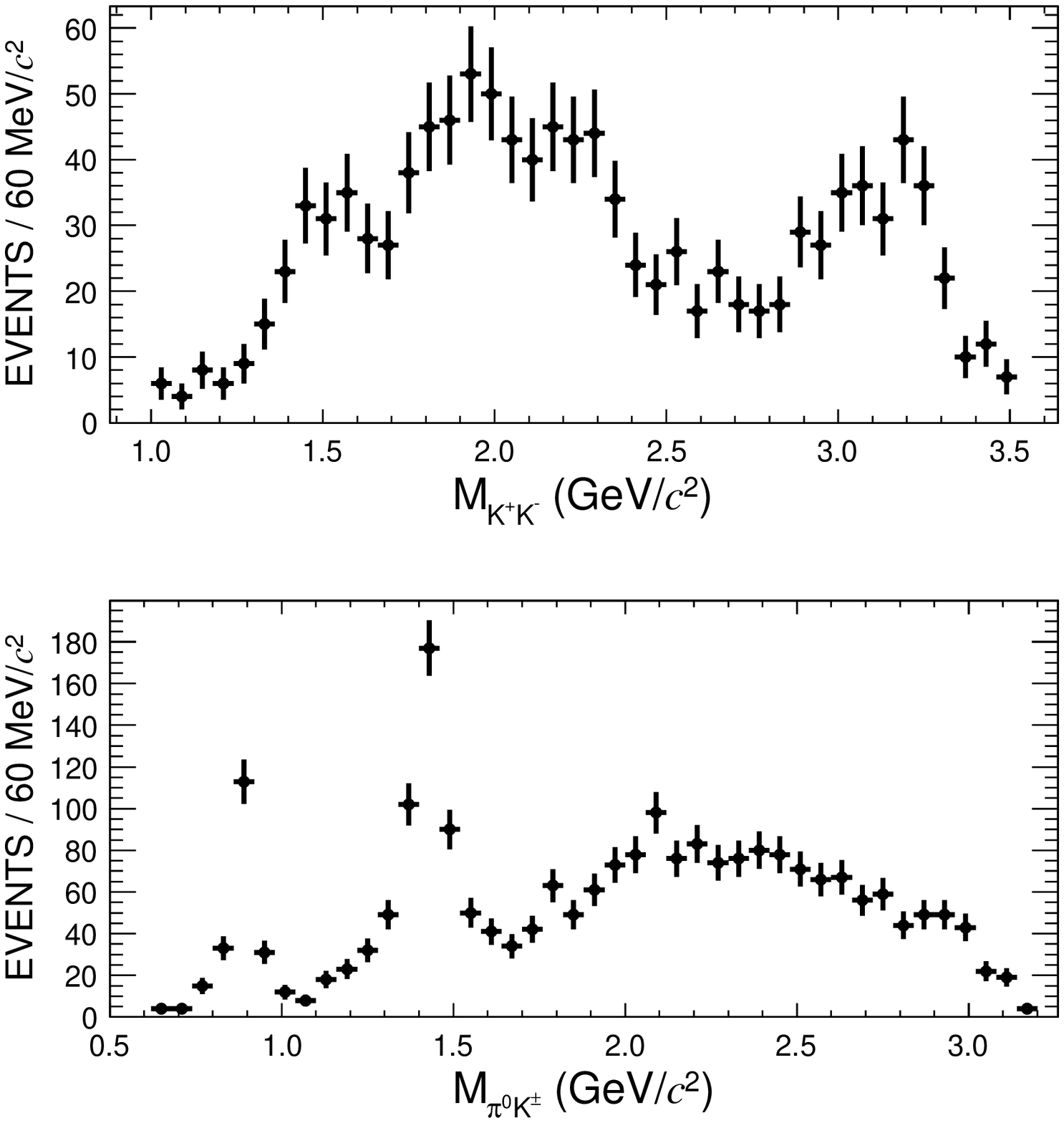}
   \put(-230,240){(a)}  \put(-230,100){(b)}
  \caption{The invariant mass projection of the Dalitz plot (see Fig. \ref{MKpi_MKK}) for the $\psip\to\kk\pi^0$ decay. (a) $M_{\kk}$ is plotted with one entry per event, and  (b) $M_{\pi^{0}K^{\pm}}$ is plotted with two entries per event. }
  \label{MKpi_MKK}
\end{figure}

There are in total 1158 $\psip\to\kk\pi^0$ events selected from the
data. A Dalitz plot of these events is shown in
Fig. \ref{dalitz}. Invariant mass spectra of $\pi^{0}K^{\pm}$ and
$\kk$ are shown in Fig. \ref{MKpi_MKK}. The two peaks in the $\pi^0
K^\pm$ mass spectrum correspond to the $K^{*}(892)^\pm$ and
$K^{*\pm}_J$, where $K^*_J$ may be $K^{*}_J(1430)$ or $K^{*}(1410)$. A
partial wave analysis (PWA), described below, is used to study the
Dalitz plot structures.

\subsubsection{\boldmath Final selection of $\psip\to\kk\eta$}
\begin{figure}
 \centering
 \includegraphics[width=10cm]{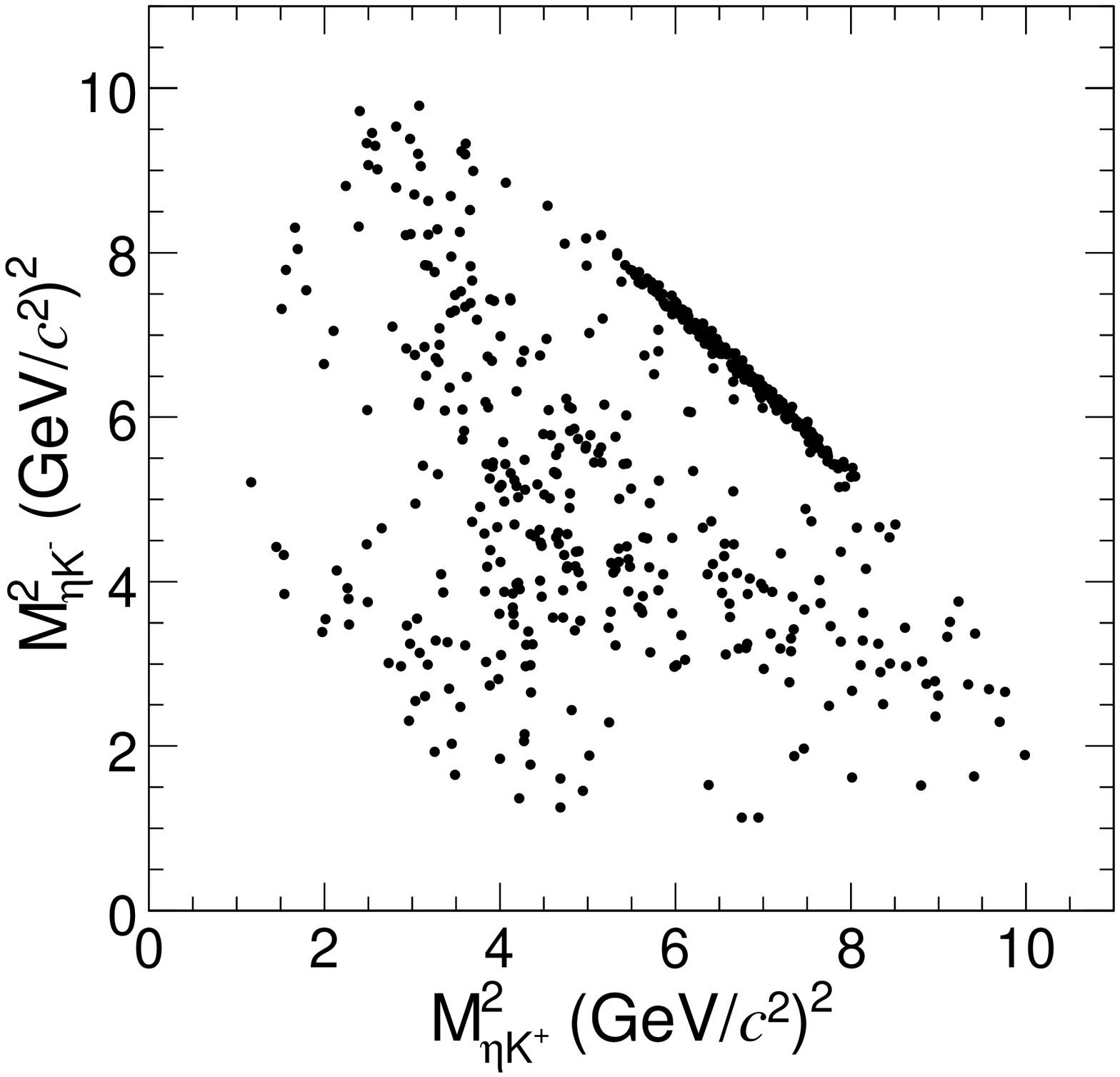}
 \caption{The Dalitz plot for $\psip\to\eta\kk$. }
 \label{dalitiz_etakk}
\end{figure}

The $\eta$ candidates are reconstructed using the two selected photons
in $\gamma\gamma\kk$, and the $\eta$ yields are
determined by a fit to the $M_{\gg}$ distribution. To suppress the
background from $\psip\to\eta\jp$, with $\jp\to\kk$, the invariant
mass of the two kaons is required to be less than 3.05 GeV$/c^2$. The
background from the decay $\psip\to\gamma\chi_{c0/2}$, with
$\chi_{c0/2}\to\pi^0/\eta\kk$ ($\chi_{c1}\to\pi^0\kk$ or $\eta\kk$ is
forbidden), is suppressed by requiring that the lower energy photon
should be outside of the range 115 MeV to 185 MeV. A Dalitz plot of
the surviving events is shown in Fig. \ref{dalitiz_etakk}, which is
produced by using a loose $\eta$ mass requirement of $0.48
~\textrm{GeV/c}^2\le M_{\gg}\le0.6 ~\textrm{GeV}/c^2$ compared to the
mass resolution for $\eta\to\gg$ ($\sim$7 MeV$/c^2$). The diagonal
band shows a clean signal for $\psip\to\phi\eta$ decays.

\section{\boldmath Partial wave analysis of $\psip\to\kk\pi^0$\label{kkpipwa}}
We perform a partial wave analysis of the decay $\psip\to\kk\pi^0$ in order to determine branching fractions for $\psip\to K^{*}(892)^\pm K^{\mp}$ and $K^{*\pm}_JK^{\mp}$.
\subsection{The method}
The method of the PWA is similar to that utilized in a previous BES
publication \cite{bespwa}. The decay amplitudes are constructed using
the relativistic covariant tensor amplitudes as described in
Ref. \cite{zoubs}. For the decay $\psip\to\kk\pi^0$, the general form
of amplitude reads:
\begin{equation}
A(m)=\psi_\mu(m)A^\mu=\psi_\mu(m)\sum_i\Lambda_i U_i^\mu,
\end{equation}
where $\psi_\mu(m)$ is the polarization vector of $\psip$ with a helicity value $m$; $U_i^\mu$ is the $i$-th partial-wave amplitude with the coupling strength determined by a complex parameter $\Lambda_i$. The differential cross section is given by
\begin{equation}
{d\sigma \over d\Phi}={1\over 2}\sum_{m=\pm1}A^\mu(m) A^{*\mu}(m)=\sum_{m,i,j}P_{ij}\cdot F_{ij},
\end{equation}
where $P_{ij}= P^{*}_{ji}\equiv \Lambda^i\Lambda^{*j}$ and $F_{ij}=F_{ji}^*\equiv {1\over 2}\sum_{\mu=1}^2U_i^\mu U_j^{*\mu}$. Here, the sum over the $\psip$ polarization is taken as $m=\pm1$ since the $\psip$ particle is produced from $\ee$ annihilation. The partial wave amplitudes $U_i$ for the intermediate states, e.g. $K^{*}(892)^\pm K^{\mp}$, $K_2^{*}(1430)^\pm K^{\mp}$ etc., are constructed from the $K^+,~K^-$ and $\pi^0$ four-momenta. In the amplitude, the line shape for the resonance is described with a Breit-Wigner function:
\begin{equation}
BW(s)={1\over M^2-s-iM\Gamma},
\end{equation}
where $s$ is the invariant-mass squared, and $M$ and $\Gamma$ represent the mass and width, respectively.

The relative magnitudes and phases for amplitudes $U_i$ are determined
by an unbinned maximum likelihood fit. The joint probability density
for observing the $N$ events in the data sample is
\begin{equation}
\mathcal{L}=\prod_{i=1}^N P(x_i),
\end{equation}
where $P(x_i)$ is a probability to produce event $i$ with four-vector momentum $x_i=(p_{K^+},p_{K^-},p_{\pi^0})_i$. The normalized $P(x_i)$ is calculated from the differential cross section
\begin{equation}
P(x_i)={(d\sigma /d\Phi)_i \over \sigma_{MC}},
\end{equation}
where the normalization factor $\sigma_{MC}$ is calculated from a MC
sample with $N_{MC}$ accepted events, which are generated with a phase
space model and then subject to the detector simulation, and are
passed through the same event selection criteria as applied to the
data analysis. With an MC sample of sufficiently large size, the
$\sigma_{MC}$ is evaluated with
\begin{equation}
\sigma_{MC}={1\over N_{MC}}\sum_{i=1}^{N_{MC}}\left({d\sigma\over d\Phi}\right)_i.
\end{equation}
For technical reasons, rather than maximizing $\mathcal{L}$, $S=-\ln\mathcal{L}$ is minimized using the package FUMILI \cite{fumili}.
\subsection{Background subtraction}
The number of non-$\pi^0$ background events in the selected $\kk\pi^0$ data sample, estimated from a $\pi^0$ sideband defined by $M_{\gg}\in [0.079,0.109]$ and $[0.165,0.195]$ GeV$/c^2$, is 43$\pm$7 events.
The MC simulation shows that these background events are mainly due to $\psip\to\gamma\chicj,~\chicj\to\gamma\kk$ or $~\pi^0\kk$. A low level of non-$\kk$ background (3 events) comes from $\psip\to\pi^0\pi^0\jp,~\jp\to\mu^+\mu^-$ due to a misidentification of muons as kaons.

Events from the QED process, $\ee\to\gamma^*\to \kk\pi^0$ produced at
a center-of-mass energy corresponding to the mass of the $\psip$ peak,
have the same final state as our signals of interest. Background from
this source is estimated from two data sets taken at $\sqrt s=$ 3.773
GeV and 3.65 GeV.  Since the decay of $\psi(3770)\to \kkp$ is not
observed \cite{pdg}, the events obtained at $\sqrt s=3.773$ GeV are
regarded as all due to the QED process.  After normalizing their
integrated luminosities to that of the $\psip$ sample, the number of
events obtained at each of the data sets are 195$\pm$3 and 195$\pm$27,
respectively, and in good agreement with each other.

\begin{figure}
 \centering
 \includegraphics[width=10cm]{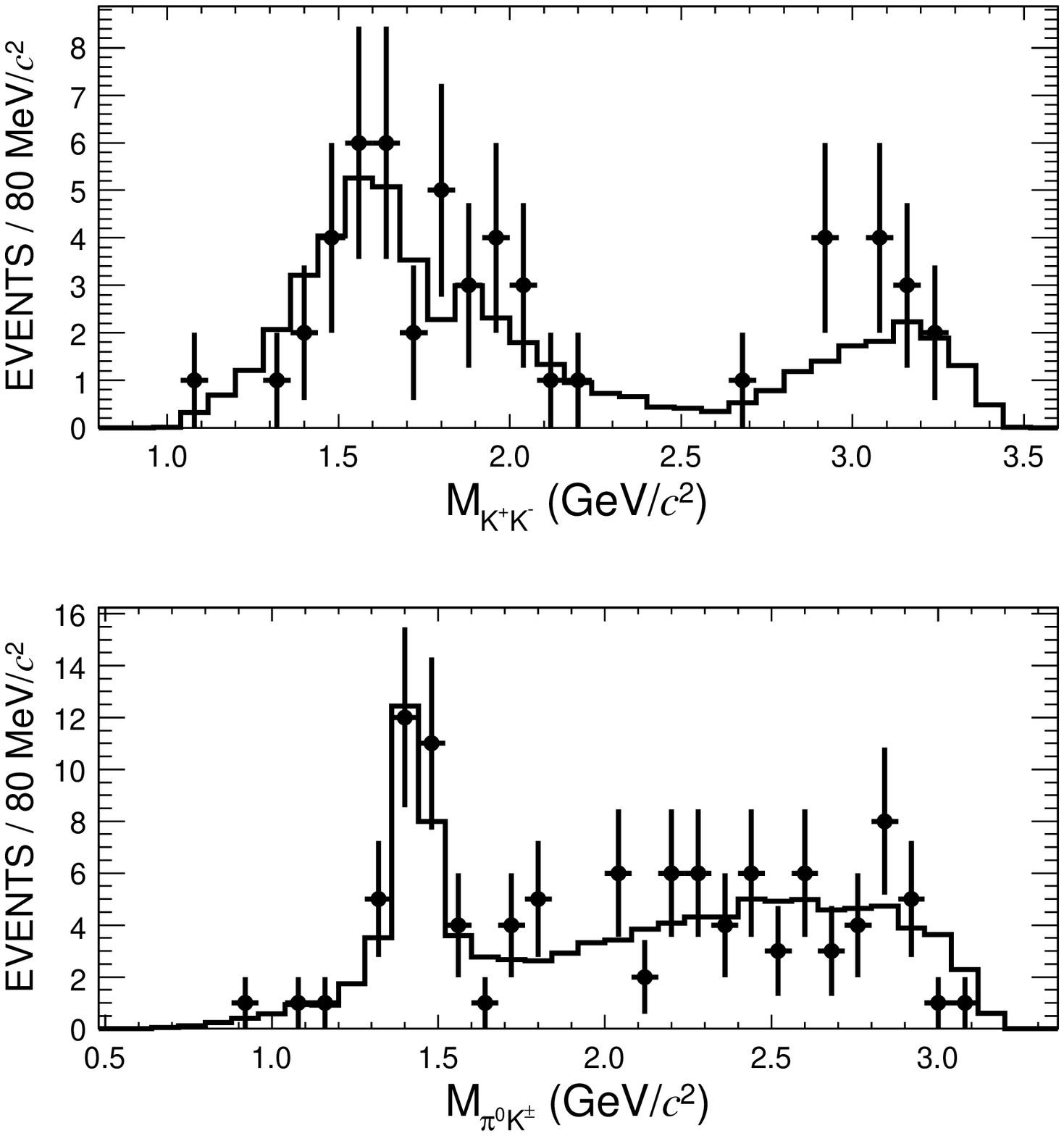}
 \put(-230,240){(a)}  \put(-230,100){(b)}
 \caption{The $\kk$ (one entry per event) and $K\pi$ (two entries per event) invariant mass distributions at $\sqrt s=3.65$ GeV. The dots with error bars are data and the histograms are MC events as described in the text. }
 \label{pi0KK:PWA_3650_M}
\end{figure}

The QED background events at the $\psip$ peak are generated using a model determined by performing a PWA fit to the data set taken at 3.773 GeV. As a cross check, the model with the determined coupling strengths is used to generate MC samples and compared with the data set taken at 3.650
GeV. Figure~\ref{pi0KK:PWA_3650_M} compares mass distributions obtained from MC events with those obtained from experimental data. Here MC and experimental
data were generated or taken at $\sqrt s=3.650$ GeV.
For the $\kk$ and $K\pi^0$ invariant mass distributions, the data and MC agree well within statistical errors, and a peak around
$M_{\pi^0 K^\pm}=1.4$ GeV$/c^2$ can be seen.

In the PWA fit, background events obtained from MC simulation or $\pi^0$ mass sideband are used to account for the background events in the data using a negative log-likelihood value. Hence, the complete log-likelihood function is:
\begin{equation}
\ln \mathcal{L}=\ln \mathcal{L}_{dt} - \sum\ln\mathcal{L}_{bg},
\end{equation}
where $\mathcal{L}_{dt}$ and $\mathcal{L}_{bg}$ are the likelihoods determined with the data and background events, respectively.
The backgrounds are divided into two kinds: reducible background and irreducible background (QED background).
This technique of background treatment assumes no interference between signal and irreducible background events.
This method has been used in the analysis of Crystal Barrel data~\cite{crstal} and  BESII data~\cite{bespwa,pwachi0}.

\subsection{Analysis results}
Motivated by the structures seen in the Dalitz plot (Fig.~\ref{dalitz}) and its projections (Fig.~\ref{MKpi_MKK}), the decay modes listed in Tables \ref{PWAmode} and \ref{additionalRes} are considered in the PWA fit. Only the modes with a statistical significance larger than 5 standard deviation ($\sigma$) are taken as the best solution, which includes the resonances $K^*(892)^\pm,~K^*_2(1430)^\pm,~K^*(1680)^\pm$ and $\rho(1700)$, and the non-resonance mode $\kkp$ (see Table \ref{PWAmode}). The significance of a mode is calculated by comparing the difference of the $S(=-\ln \mathcal{L})$ values between the fit with and without that mode. The non-resonance mode is described as a $P-$wave $\kk$ system. For the charge-conjugate channels, the coupling strengths in amplitudes are the same. Each mode in the amplitude introduces two parameters are determined by the PWA fit, the magnitude of the coupling strength and the phase angle.

Other intermediate states, like $\rho(770)$, $\rho(1450)$, $\rho(1900)$, $\rho(2150)$ in the $\kk$ final states, and $K^{*}(1410)^\pm$ and $K^{*}(1980)^\pm$ in the $\pi^0 K^{\pm}$ final states, were considered and tested in the PWA fit. Adding them to the best solution does improve the fit quality, but these additional modes have a statistical significance of less than 5$\sigma$ (see Table \ref{additionalRes}). The $\rho(770)$ may decay to $\kk$ if its mass is larger than $\kk$ threshold, but its significance is 4.6$\sigma$.
A $P-$wave $\pi^0K^\pm$ system as an additional non-resonance contribution was tried and had a significance of 1.9$\sigma$. The variations
to the $K^*(892)^\pm$ and $K^*_2(1430)^{\pm}$ signal yields by including these intermediate states are included as a systematic uncertainty.

\begin{table}[htbp]
 \centering
 \caption{The significance and number of events of each resonance under the best solution.}
 \label{PWAmode}
 \begin{tabular}{cccc}
 \hline\hline
 Decay                       &Fitted events     & Significance($\sigma$)\\ \hline
 $K^{*}(892)^{\pm}K^{\mp}$        &224$\pm$21        &26.5\\
 $K^{*}_{2}(1430)^{\pm}K^{\mp}$   &251$\pm$22        &21.0\\
 $K^{*}(1680)^{\pm}K^{\mp}$       &115$\pm$20        &11.1\\
 $\rho^0(1700)\pi^0$              &59$\pm$10         &8.7 \\
 $\kkp$                           &721$\pm$60        &18.8\\\hline\hline
 \end{tabular}
\end{table}

\begin{table}[htbp]
 \centering
 \caption{Significance for additional resonance.}
 \label{additionalRes}
 \begin{tabular}{cc}\hline\hline
  Decay                      &Significance($\sigma$)\\ \hline
  $\rho^0(770)\pi^0$              &4.63\\
  $\rho^0(1450)\pi^0$             &4.40\\
  $\rho^0(1900)\pi^0$             &1.13\\
  $\rho^0(2150)\pi^0$             &3.21\\
  $\rho^0_3(1690)\pi^0$           &1.84\\
  $K^{*}(1410)^{\pm}K^{\mp}$      &2.23\\
  $K^{*}_2(1980)^{\pm}K^{\mp}$    &2.14\\
  $K^*_3(1780)^{\pm}K^{\mp}$      &3.05\\
  $K^*(2045)^{\pm}K^{\mp}$        &3.26\\
  non-resonance ($K^{\mp}\pi^0$)   &1.89\\
  \hline\hline
  \end{tabular}
\end{table}

For intermediate states around $K\pi$ invariant mass of 1.43 GeV, there are four established resonances, namely,
$K_1(1400)$, $K^*(1410)$, $K_0^*(1430)$ and $K_2^*(1430)$; according to the spin-parity conservation, only $K^*_2(1430)$ and $K^*(1410)$ are
allowed. If $K_2^*(1430)^\pm K^\mp$, which is the best solution in the PWA, is replaced with $K^*(1410)^\pm K^\mp$, the fit
fails to match the data,
and the log-likelihood gets worse by 126, and the contribution from the $K^*(1410)$ is negligible. If
$K^*(1410)^\pm K^\mp$ is taken in addition to $K^*_2(1430)^\pm K^\mp$ to the best solution, the log-likelihood only improves by 3.65, corresponding to a significance of 2.2$\sigma$.

The non-resonance decay $\psip\to\kk\pi^0$ is indispensable in the
fit, with a statistical significance of $19\sigma$. We have tried to
replace it with a broad resonance, such as $\rho(2150)\pi^0$. The fit
fails to match the data, and the log-likelihood gets worse by 95. Note
that the total number of fitted events $1370\pm70$ in
Table~\ref{PWAmode} is larger than the number of net $\kkp$ events
917(=1158-241) due to the destructive interference among the included
resonances.

The numbers of fitted events given in Table \ref{PWAmode} are derived
from numerical integration of the resultant amplitudes as done in
Ref. \cite{pwachi0}. The statistical errors are derived from the $S$
distribution versus the number of fitted events; one standard
deviation corresponds to the interval that produce a change of
log-likelihood of 0.5. When performing the PWA fit to the data, the
masses and widths of the intermediate states are fixed at the PDG
values, and their errors quoted in the PDG are used to estimate the
associated systematic errors.

\begin{figure}
  \centering
  \includegraphics[width=10cm]{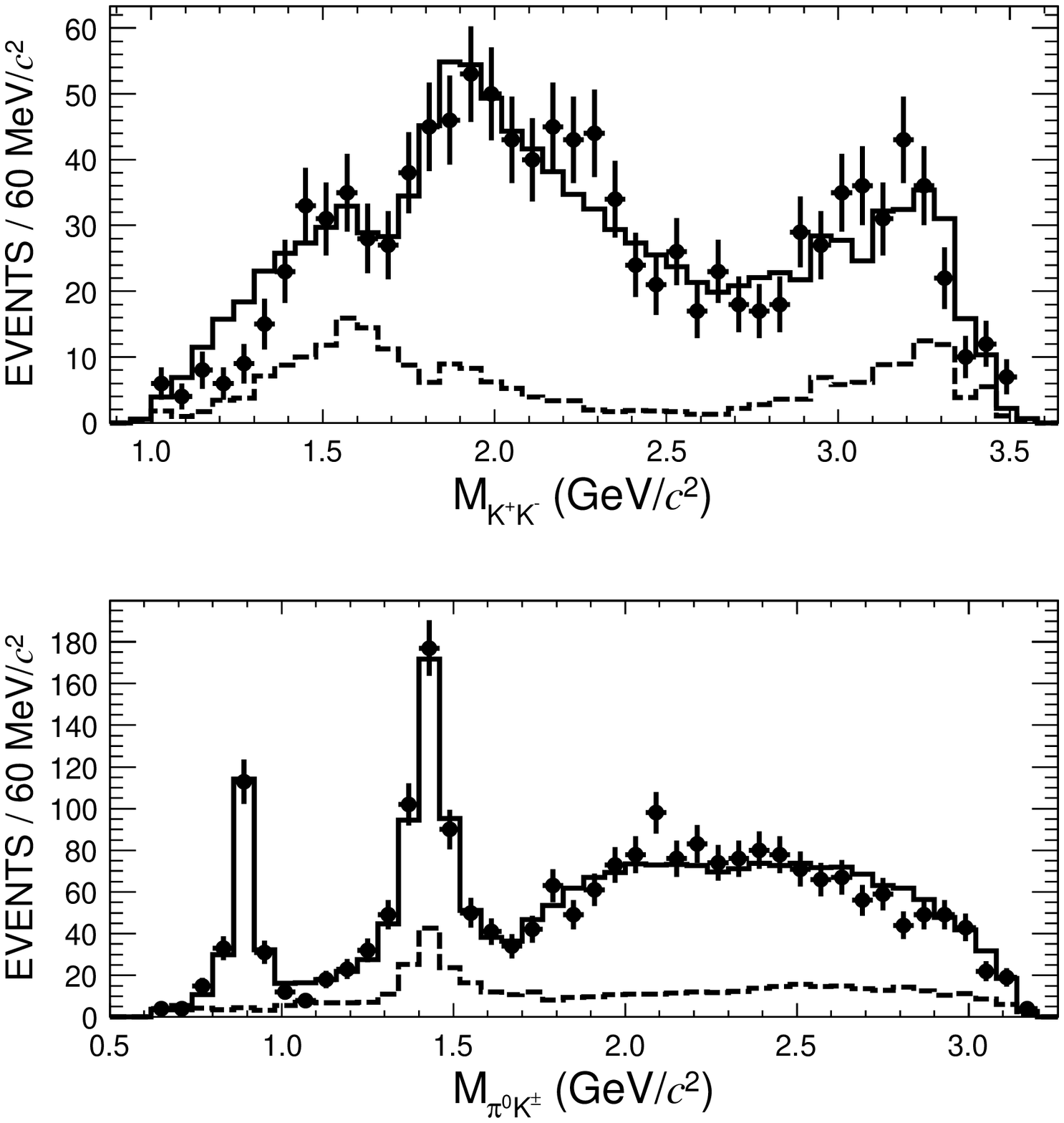}
  \put(-230,240){(a)}  \put(-230,100){(b)}
  \caption{The results of fit to (a) $\kk$ and (b) $\pi^0 K^{\pm}$ mass distributions for the data, where points with error bars are data and histograms are total fit results. The dashed histograms are the sum of the background sources, including QED and non-$\kk\pi^0$ contributions.}
  \label{fitMass}
\end{figure}

\begin{figure}
  \centering
  \includegraphics[width=10cm]{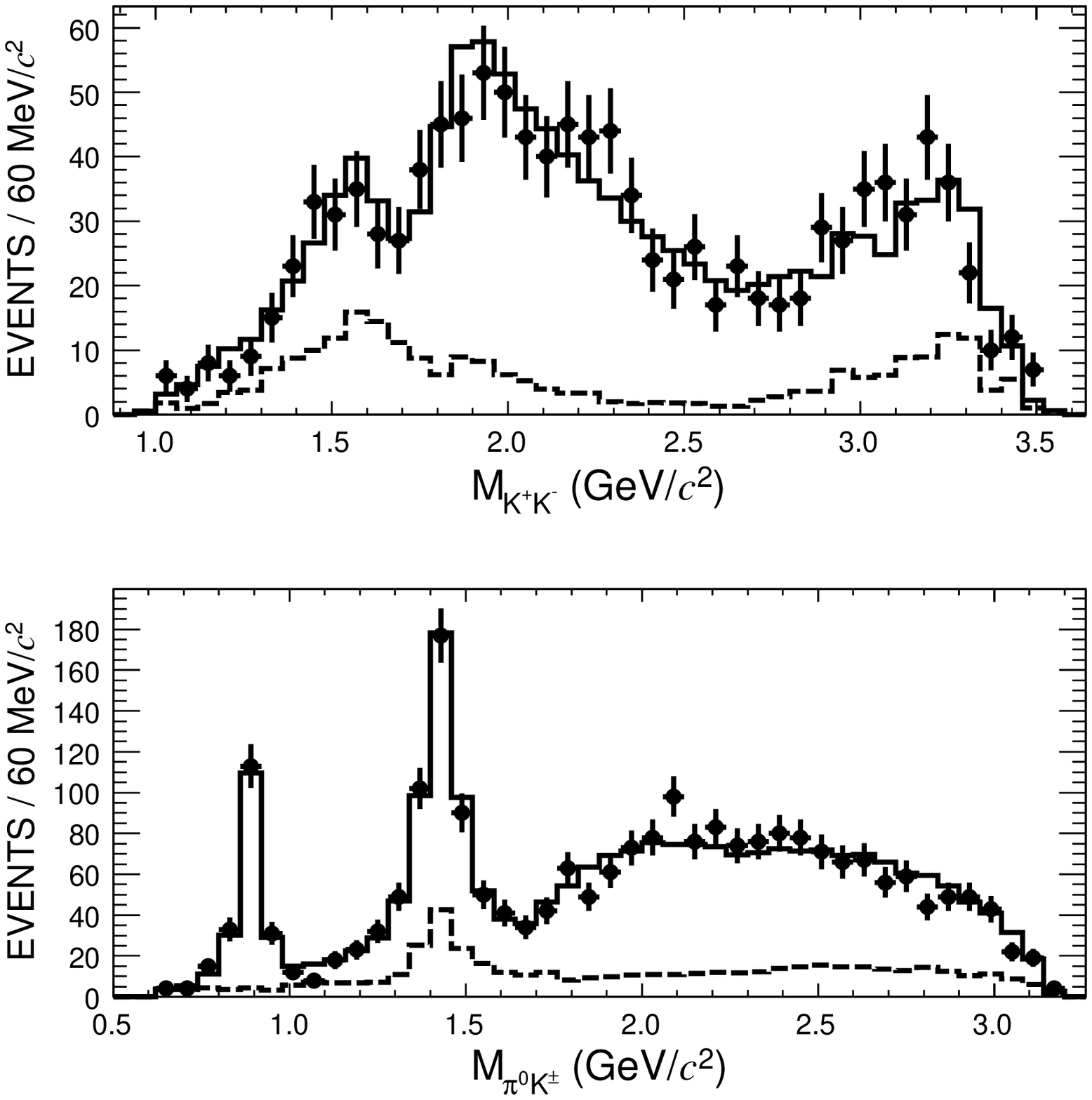}
  \put(-230,240){(a)}  \put(-230,100){(b)}
  \caption{The results of fit to (a) $\kk$ and (b) $\pi^0 K^{\pm}$ mass distributions for the data; where points with error bars are data; histograms denote total fit results with an additional mode of $\rho(1450)\pi^0$ being added to the best solution of the PWA fit (see Fig.~\ref{fitMass}). The dashed histograms are the sum of the background sources, including QED and non-$\kk\pi^0$ contributions.}
  \label{PWA_rho1450}
\end{figure}

Figure \ref{fitMass} depicts a comparison between the data and the
best solution obtained from the PWA fit to the data. Here the
projected $M_{\kk}$ and $M_{\pi^0K^\pm}$ mass distributions are
shown. They are in general in a good agreement except for several
points at the low $M_{\kk}$ mass region. An additional
$\rho(1450)\pi^0$ to the best solution in the PWA helps to
improve the fit quality through destructive interference (see
Fig. \ref{PWA_rho1450}). The statistical significance of this
additional mode is only about 3.2$\sigma$ and it only brings a small
difference in signal yields, 3.3\% for $K^{*}(892)^\pm K^{\mp}$
and 0.4\% for $K_2(1430)^{*\pm} K^{\mp}$. These yield differences
are taken as a systematic uncertainties to account for additional
resonance contributions to the low $M_{\kk}$ mass region.

The goodness of the global fit is determined by calculating a
$\chi^2_{\textrm{all}}$ defined by
\begin{equation}
\chi^2_{\textrm{all}} = \sum_{j=1}^{5}\chi^2_j \textrm{,~with~}\chi^2_j=\sum_{i=1}^N{(N_{ji}^{DT}-N_{ji}^{Fit})^2\over N_{ji}^{Fit}},
\end{equation}
where $N_{ji}^{DT}$ and $N_{ji}^{Fit}$ are the number of events in the $i$-th bin for the distribution of the $j$-th kinematic variable. If the measured values $N_{ji}^{DT}$ are sufficiently large, then the  $\chi^2_{\textrm{all}}$ statistic follows the $\chi^2$ distribution function with the number of degrees of freedom (ndf) equal to the total bins of histograms  \footnote{In a histogram, bins with event entries less than 10 are combined as one bin.} minus the number of fitted parameters; and the individual $\chi^2_j$ gives a qualitative measure of the goodness of the fit for each kinematic variable.

\begin{figure}
  \centering
  \includegraphics[width=10cm]{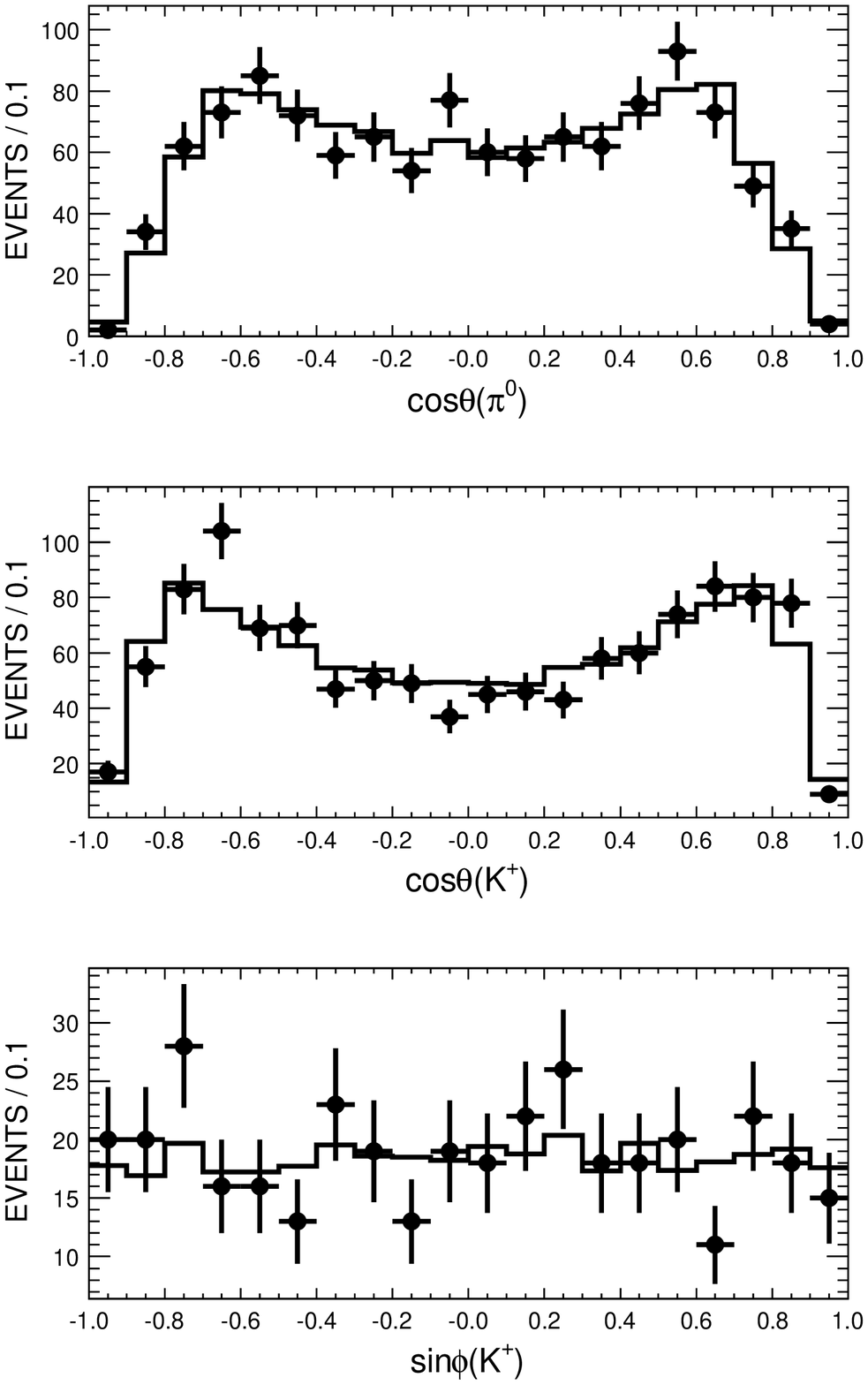}
  \put(-120,380){(a)}  \put(-120,230){(b)}
  \put(-120,85){(c)}
  \caption{The fit results of the angular distributions, where the points with error bars are data and histograms are the fit results. (a) $\cos\theta$ distribution for $\pi^0$, (b) $\cos\theta$ distribution for $K^+$, and (c) $\sin\phi$ distribution for $K^+$, here the angles are defined in the $\psip$ rest frame.}
  \label{fitAng}
\end{figure}

For the 3-body decay $\psip\to\kk\pi^0$, there are 5-independent variables, which are selected as the mass of the $\kk$ system ($M_{\kk}$), the mass of
the $\pi^0 K^{\pm}$ system ($M_{\pi^0 K^{\pm}}$), the polar angle for the $\pi^0$ ($\theta_{\pi^0}$), the polar angle for the $K^-$ ($\theta_{K^-}$),
and the azimuthal angle for the $K^+$ ($\phi_{K^+}$), where the angles are defined in the $\psip$ rest frame.  Figure \ref{fitAng} compares the angular
distributions between the best fit solution and the data, and a good agreement can be observed. A sum of all these $\chi^2_j$ values gives $\chi^2_{\text{all}}=147.70$, and the total number of degrees of freedom (126) is taken as the sum of the total number of bins having non-zero events minus the total number of parameters in the PWA fit. The global fit goodness $\chi^2_{\textrm{all}}/ndf$ is 1.2.

\subsection{Branching fractions}
Branching fractions for $\psip\to K^{*}(892)^+K^{-}+c.c.$, $\psip\to K^{*}_2(1430)^+K^{-}+c.c.$, and the inclusive decay
$\psip\to\kk\pi^0$ (including all resonances) are calculated
\begin{eqnarray}
\nonumber
Br(\psip\to K^{*+}K^{-}+c.c.)&=&{N^{obs}_{K^*}\over \varepsilon N_{\psip} Br(K^{*+}\to K^{+}\pi^0)Br(\pi^0\to \gamma\gamma)},\\
Br(\psip\to \kk\pi^0)& =& {N^{obs}_{\kkp}\over \varepsilon N_{\psip} Br(\pi^0\to \gamma\gamma)}.
\end{eqnarray}
Here $Br(K^{*+}\to K^{+}\pi^0)$ is the branching fraction for
$K^*(892)^{+}$ (33.23\%) or $K_2^*(1430)^{+}$ (16.60\%) resonances;
$N^{obs}_{K^*}$ is the signal yield obtained from the PWA fit
($224\pm21$ and $251\pm22$ for $K^*(892)$ and $K_2^*(1430)$,
respectively); $N_{\kkp}^{obs}$ is the net number of $\kkp$ events
($917\pm37$); $N_{\psip}=(106\pm4)\times 10^{6}$ is the number of
$\psip$ events\cite{psiprimdecays}; and $\epsilon$ is the detection
efficiency. To determine $\epsilon$, the intensity from the amplitudes
is used to weight both the complete set of generated MC events and the
set which survives the selection procedure, and the ratio between
these two weighted sets is taken as the detection efficiency.

The branching fractions are measured to be:
\begin{eqnarray}
Br(\psip\to\pi^0\kk)&=&(4.07\pm0.16)\times 10^{-5},\\
Br(\psip\to K^{*}(892)^{+}K^{-}+c.c.)&=&(3.18\times 0.30)\times 10^{-5},\\
Br(\psip\to K^*_2(1430)^+K^{-}+c.c.) &=& (7.12\pm0.62)\times 10^{-5},
\end{eqnarray}
where the errors are only statistical.

\section{\boldmath $\psip\to\pi^0\phi$}
\begin{figure}
  \centering
  \includegraphics[width=10cm]{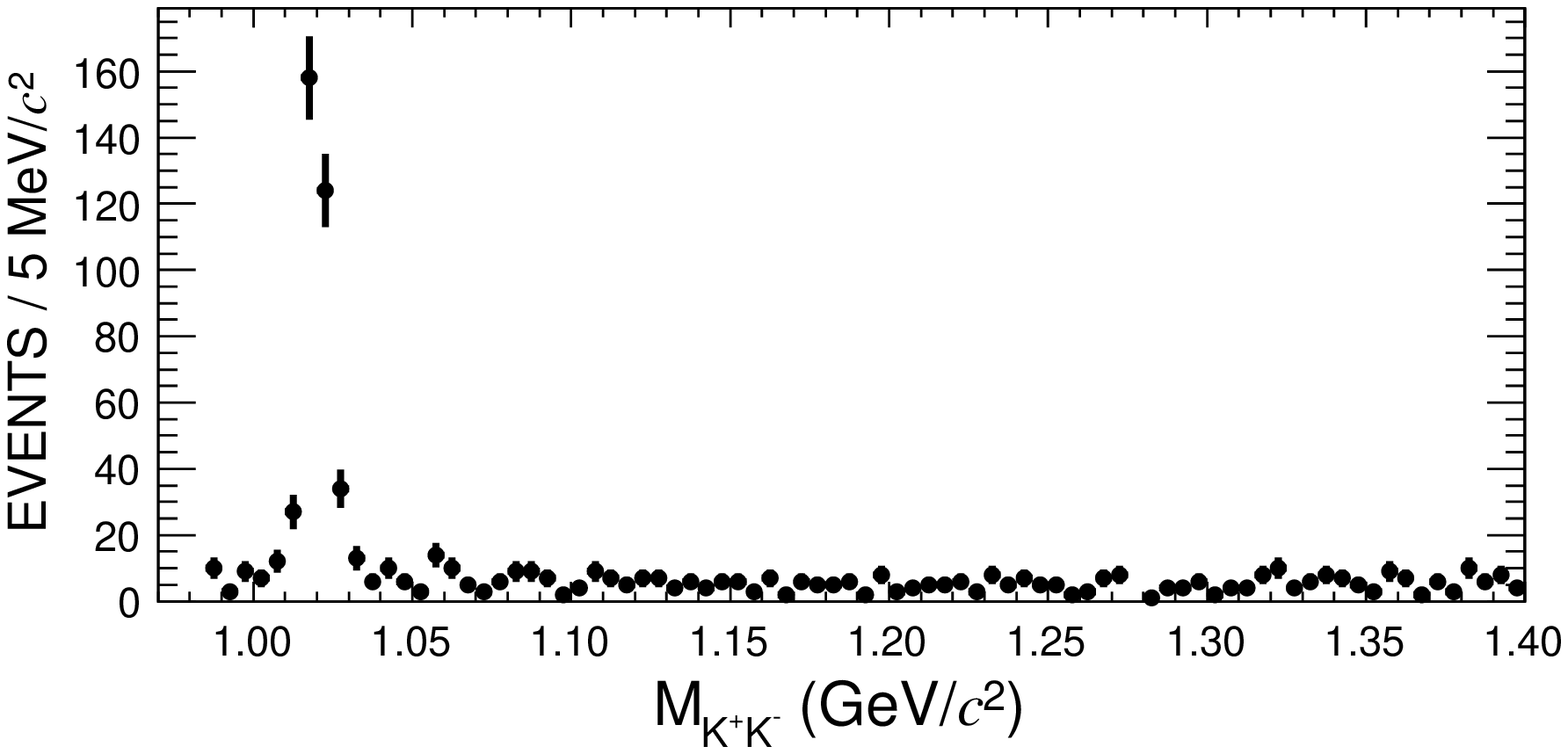}
  \caption{The $\kk$ invariant mass  selected in $\psip\to\gg\kk$.}
  \label{phi_sig}
\end{figure}

\begin{figure}
  \centering
  \includegraphics[width=10cm]{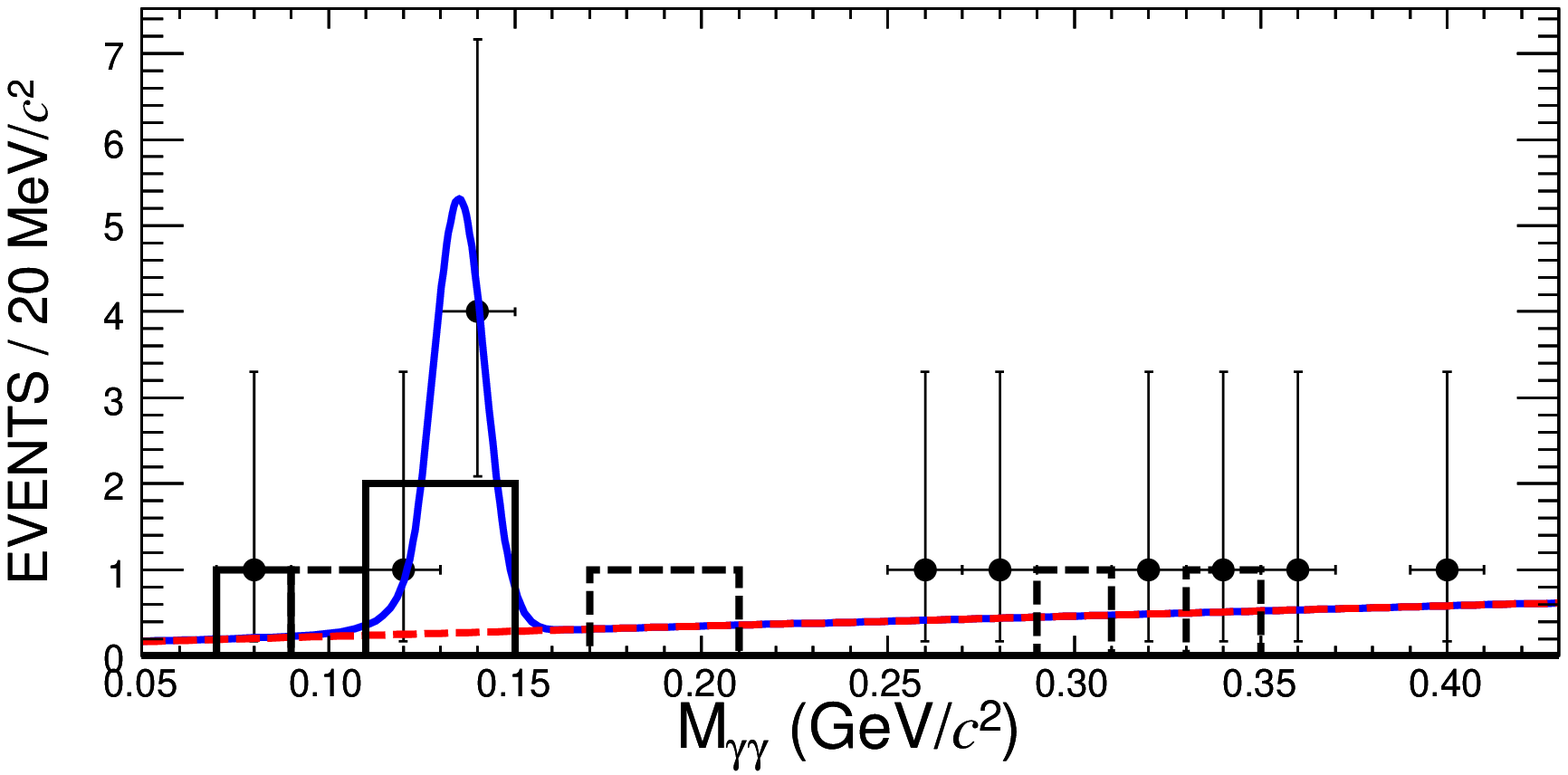}
  \caption{(Color online) The invariant mass distribution of two photons in the selected $\psip\to\phi\gg$ events; the solid line shows a fit to $\pi^0$; the dashed line shows the fitted background and comparison to the backgrounds estimated with $\phi$ sideband (line histogram) and MC simulation (dashed histogram).}
  \label{mgg_for_phipi0}
\end{figure}

The $\phi$ candidates for $\psip\to\phi\pi^0$ are reconstructed using the two kaons selected in the decay $\psip\to\kk\gg$. Figure \ref{phi_sig} shows the invariant mass distribution of the two kaons, and a $\phi$ signal is clearly seen. The $\phi$ candidates are selected by requiring $|M_{\kk}-m_{\phi}|< 10$ MeV$/c^2$, where $M_{\kk}$ and $m_{\phi}$ are the invariant mass of the two kaons and the mass of the $\phi$ \cite{pdg}. Background sources from the initial state radiation process $\ee\to\gamma\phi$ are suppressed by requiring that the energy for the energetic photon is less than
1.6 GeV. Figure \ref{mgg_for_phipi0} shows the invariant mass distribution of the two photons after the $\phi$ selection criterion is applied. No significant $\pi^0$ signal
is observed.

The number of observed events for $\psip\to\pi^0\phi$ is obtained by fitting the mass distribution of the two photons as shown in Fig.
\ref{mgg_for_phipi0}. The line shape of $\pi^0$ is taken from the MC simulation, and the background shape is taken as a first-order Chebychev
polynomial function. The fit results are shown in Fig. \ref{mgg_for_phipi0} and the significance of $\pi^0$ signal is less than 3.0$\sigma$.
The upper limit of observed $\pi^0$ events is estimated using the Bayesian approach to be $N^{up}=6$ at the 90\% confidence level.

The upper limit on the branching fraction for $\psip\to\pi^0\phi$ is calculated with
\begin{eqnarray}
 Br(\psip\to \pi^0\phi) < \frac{N^{up}}{\varepsilon N_{\psip} Br(\pi^0\to\gg) Br(\phi\to\kk)(1-\sigma^{sys})},
\end{eqnarray}
where $Br(\pi^0\to\gg)$ and $Br(\phi\to\kk)$ are the branching fractions for  $\pi^0\to\gamma\gamma$ and $\phi\to \kk$, respectively;
$N_{\psip}=(106\pm4)\times10^6$ is
the number of total $\psip$ decays; $\varepsilon=35.63\%$ is the detection efficiency that was determined using MC events generated with the angular distribution $1+\cos^2\theta$ for $\psip\to \pi^0\phi$, where $\theta$ is the $\phi$ polar angle. $\sigma^{sys}=5.8\%$ is the systematic error as listed in Table \ref{sys:gen_sys}. The upper limit of the branching
fraction is $Br(\psip\to\phi\pi^0)< 4.0\times 10^{-7}$ at the 90\% C.L.

\section{\boldmath $\psip\to\eta\kk$}
\subsection{Background analysis}
Background sources for $\psip\to\eta\kk$ are studied with the $\psip$
inclusive MC sample. The dominant background comes from
$\psip\to\gamma\gamma_{FSR}\kk$, where $\gamma_{FSR}$ is a final-state
radiation photon, $\psip\to\gamma\chi_{c2}$, with
$\chi_{c2}\to\kk\pi^0$ and $~\kk\eta$. The MC simulation shows that
the $M_{\gg}$ mass distribution of sum of these events in the region
of the $\eta$ meson is a smooth and well modeled with a polynomial
function.

Background events from QED processes are studied using events taken at $\sqrt s=3.773 $ GeV that are selected with the same criteria applied to the $\psip$ data.
The signal yields are extracted with the same fit procedure used for the $\psip$ data.
For $\eta\phi$, the contribution from the resonance decay $\psi(3770)\to\eta\phi$ is estimated to be $450\pm112$ events
using the measured cross section $\sigma=2.4\pm 0.6 $ pb \cite{psiptoetaphi}. After subtracting the resonance decays, the QED yield for
the $\ee\to\eta\phi$ at $\sqrt s=3.773$ GeV is determined to be 268$\pm$115 events. For $\eta\kk$, the observed events are considered to be
exclusively from QED processes because the $\psi(3770)\to\eta\kk$ has not observed \cite{pdg}. At the $\psip$ peak, the QED background sources are
estimated to be 16$\pm$7 events for the $\eta\phi$ and 4$\pm$1 events for the $\eta\kk$ according to the luminosity normalization. As a cross
check, we use the data taken at $\sqrt s=3.65$ GeV to determine a QED background of $25\pm9$ events. The difference between the two estimates is taken as a background uncertainty and included
into systematic errors.

\subsection{Fit results}
We performed a two-dimensional unbinned fit to the scatter plot of $M_{\kk}$ versus $M_{\gg}$ distribution assuming that $M_{\kk}$ and $M_{\gg}$ are independent variables. Motivated by the structures seen in the $M_{\kk}$
distribution, resonances including $\phi(1020),~\phi_3(1850)$ and $\phi(2170)$ are added to the fit. The fit function includes the line shapes
describing the two-body decays $\eta\phi(1020)$, $\eta\phi_3(1850)$, $\eta\phi(2170)$, the non-resonant decay $\eta\kk$, and the background. The
$\eta$ line shape is obtained from a MC simulation; the line shapes for the $\phi(1020),~\phi_3(1850)$ and $\phi(2170)$ are described as
non-relativistic Breit-Wigner functions with their masses and widths fixed to the PDG values. The Breit-Wigner function of all the
$\phi$ states are convolved with a detector resolution function. The background shapes for the $M_{\gg}$ and the $M_{\kk}$ mass distributions
are taken as first- and second-order polynomials, respectively.

\begin{figure}[hbtp]
  \centering
  \includegraphics[width=10cm]{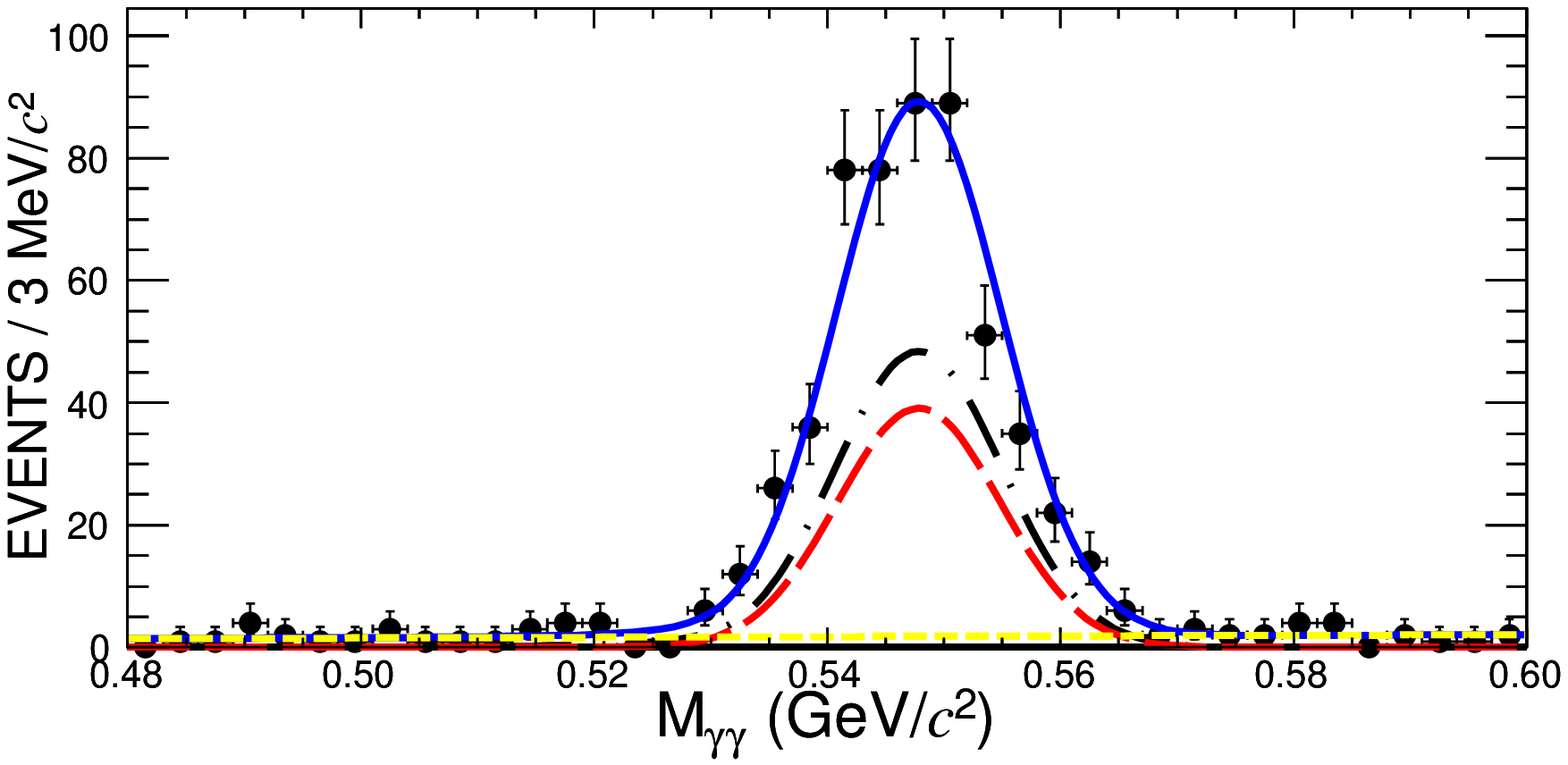}
  \caption{(Color online) Fit results projected to the two-photon invariant mass distribution $M_{\gamma\gamma}$. Dots with error bars are data. The solid line is the total fit results, and the dashed-dotted and long-dashed lines are the
  results of $\eta\phi$ and $\eta KK$ contributions, respectively. The short-dashed line is the background
  contribution.}
  \label{etakkfitMgg}
\end{figure}

\begin{figure}[hbtp]
  \centering
  \includegraphics[width=8cm]{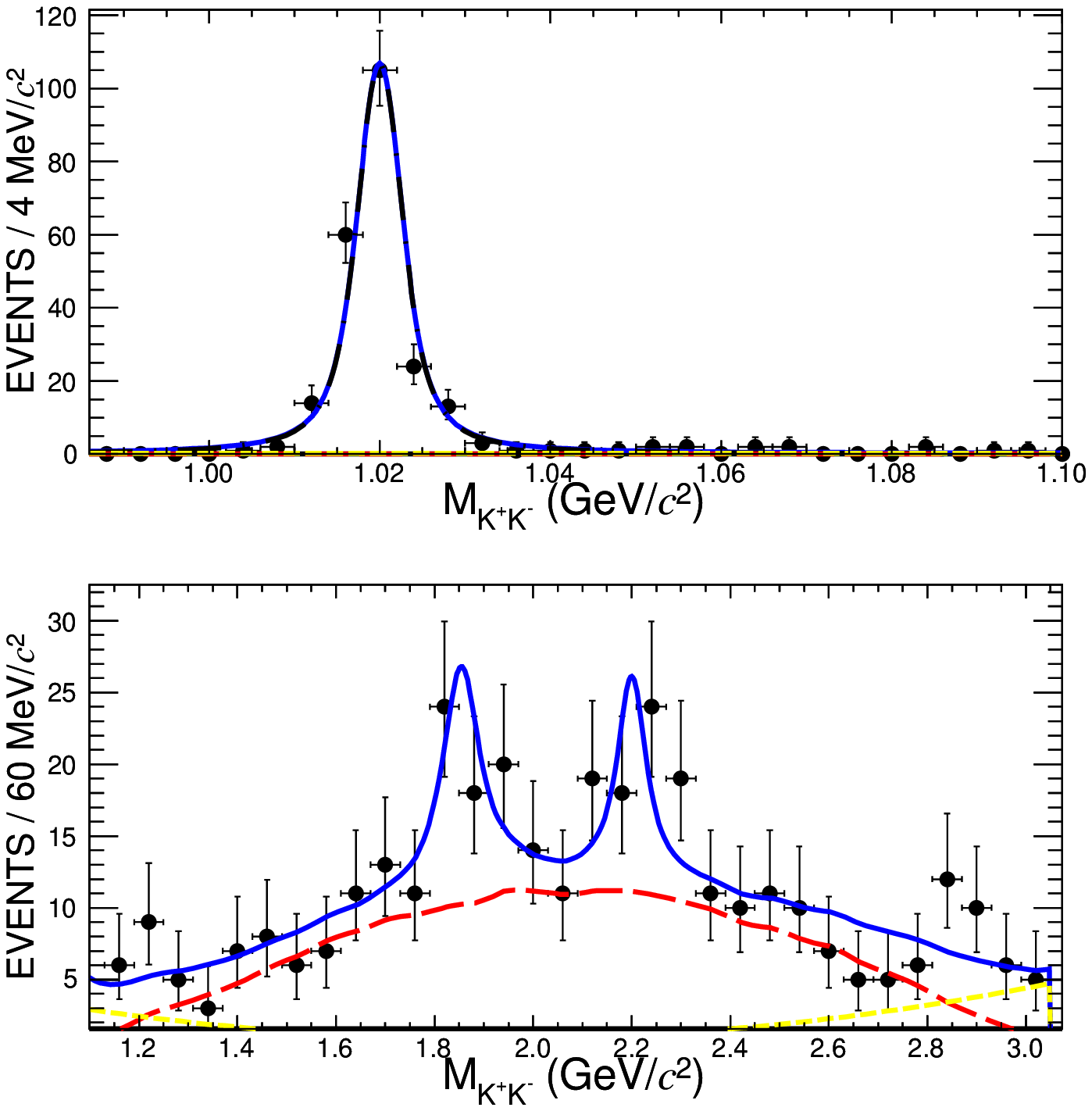}
  \put(-50,200){(a)}  \put(-50,90){(b)}
  \caption{(Color online) Fit results projected to the $\kk$ invariant mass distribution $M_{\kk}$ for (a) the  $\phi(1020)$ resonance, (b) the $\phi(1850)$ and $\phi(2170)$ resonances.  Dots with error bars are data. The solid lines are the total fit results, and the dashed-dotted and long-dashed lines are the
  results of $\eta\phi$ and $\eta \kk$, respectively. The short-dashed line is the background
  contribution.}
  \label{etakkfitMkk}
\end{figure}

The fit results after projecting to the mass distributions are shown in Figs.~\ref{etakkfitMgg} and  \ref{etakkfitMkk}.
The signal yield for the $\psip\to\eta\phi$ channel is $232\pm16$ events. Adding the $\phi_3(1850)$ and
$\phi(2170)$ resonances to the fit improves the fit quality with a statistical significance of 3.8$\sigma$ for the
$\phi_3(1850)$, and 3.1$\sigma$ for the $\phi(2170)$. The goodness of the fit is $\chi^2/ndf=0.32(0.43)$ for the
$M_{\gg}(M_{\kk})$ distribution. The yields of $\eta\phi_3(1850)$ and $\eta\phi(2170)$ plus the contribution from the non-resonance decay $\psip\to\eta\kk$
totals $288\pm27$ events. After subtracting the QED background, the net signals are $216\pm16$ events for $\psip\to \eta\phi$, and $284\pm27$ events for $\psip\to\eta\kk$.

\subsection{Branching fractions}
Branching fractions are calculated from the relations
\begin{eqnarray}
   Br(\psip\to\eta \kk)&=&\frac{N^{obs}_{\eta KK}}{\varepsilon_{\eta KK} N_{\psip}Br(\eta\to\gg)}, \\
   Br(\psip\to\eta\phi)&=&\frac{N^{obs}_{\eta\phi}}{\varepsilon_{\eta\phi} N_{\psip}Br(\eta\to\gg)Br(\phi\to\kk)}.
\end{eqnarray}
Here $N^{obs}_{\eta KK}=284\pm27$ and $N^{obs}_{\eta\phi}=216\pm16$ are the numbers of net signal events; $Br(\eta\to\gg)$ and $Br(\phi\to\kk)$ are the branching fractions for the $\eta\to\gg$ and $\phi\to\kk$ decays, respectively; $\varepsilon_{\eta KK}=22.10\%$ and $\varepsilon_{\eta\phi}=33.53\%$ are the detection efficiencies determined from MC simulations, whose angular distributions match the data; $\varepsilon_{\eta KK}$ is a weighted average for $\psip\to\eta\kk$, $\eta\phi_3(1850)$ and $\eta\phi(2170)$.
The branching fractions are calculated to be $Br(\psip\to\eta KK)=(2.97\pm0.28)\times10^{-5}$ and $Br(\psip\to\eta\phi)=(3.08\pm0.29)\times10^{-5}$, where the errors are only statistical.

\section{Systematic errors}
The systematic errors in the branching fraction measurement originated from following sources are considered:
\begin{enumerate}
\item{ photon efficiency}\\
The soft and hard photon efficiencies are studied using $\psip\to\pi^0\pi^0\jp, \jp\to\ee,\mu^+\mu^-$ and $\jp\to\rho\pi\to\pi^+\pi^-\pi^0$ decays. The difference in the photon efficiency between the MC simulation and data is 1\%, which is taken as a systematic uncertainty.

\item kaon tracking and PID efficiency\\
The uncertainties of kaon tracking and PID efficiency are studied using a sample of $\jp\to K^{*}(892)^0 K^0_S+c.c.\to K_S^0 K^+\pi^- +c.c.\to K^+\pi^-\pi^+\pi^-+c.c.$ events as done in \cite{chicj2gv}. The uncertainties for both tracking and PID are determined to be 1\% per track.

\item Number of $\psip$ events\\
The number of $\psip$ events is determined using its hadronic
decays. The uncertainty is 4\% \cite{psiprimdecays}.

\item branching fractions\\
The uncertainties of branching fractions for $K^*(892)^{\pm}/K_2(1430)^{\pm}\to K^{\pm}\pi^0,~\pi^0/\eta\to\gg$ and $\phi\to\kk$ are taken from the world average values \cite{pdg}.

\item kinematic fit\\
The differences between the MC simulation and data in the $\chi^2$ distribution of the kinematic fit arise mainly due to
inconsistences in the charged track parameters. The kaon track parameters in the MC
simulation are corrected by smearing them to match the data. The difference in the detection efficiency between with and without making a correction to
the MC is taken as a systematic error. The uncertainties are listed in Table \ref{sys:gen_sys}.

\item the $\pi^0$ mass window\\
The uncertainty due to the $\pi^0$ mass window is studied by comparing the $\pi^0$ selection efficiency obtained in the MC and the data. The uncertainty is 1.1\%.

\item fit uncertainty\\
The fit uncertainties in the $\eta\kk$ and $\eta\phi$ modes are determined by changing the fit range and background shapes. The fit range of two photons is changed to be [460, 620] MeV$/c^2$ or [470, 670] MeV/$c^2$. It is estimated to be 3.6\% (0.6\%) for $\eta\kk$ ($\eta\phi$). The background function is changed from 1st-order to 3rd-order polynomials. The uncertainties due to the background shapes are 1.6\% and 0.4\% for $\eta\kk$ and $\eta\phi$, respectively.

\item QED backgrounds \\
The QED background subtracted from $\eta\phi$ is determined with the data taken at $\sqrt s=3.773$ GeV and at $\sqrt s=3.65$ GeV. The difference in the number of QED events between these two samples is 4.5\%, which is taken as the QED background associated uncertainty.

\item additional resonances for $\eta\kk$\\
The existence of $\phi_3(1850)$ and $\phi(2170)$ intermediate states in $\eta\kk$ cannot be determined due to the
low statistics. The difference between the branching fractions determined by including and excluding these two resonances
is taken as a systematic error of 4.0\%.
\end{enumerate}
All above systematic errors are listed in Table \ref{sys:gen_sys}.

For $\kkp$, the uncertainties from the PWA fit are listed below:

\begin{enumerate}
\item Breit-Wigner form\\
The uncertainty due to the resonance line shape is evaluated by using the Breit-Wigner function with a width $\Gamma(s)$ dependent on the energy, i.e.
\begin{equation}
 \Gamma(s)=\Gamma_0{m^2\over s}\left[{p(s)\over p(m^2)}\right]^{2L+1},
\end{equation}
where $s$ is the resonance mass squared; $m$ and $\Gamma_0$ are the nominal mass and width, respectively; $p(s)$ is the magnitude of resonance momentum; $L$ is the angular momentum  for the $\psip$ decays into a two-body final state. The differences between the fit yields determined
with a constant and an energy-dependent width are taken as systematic errors. They are evaluated to be 0.1\% and 0.9\% for the $K^*(892)$ and $K^*_2(1430)$, respectively.

\item additional resonances\\
The uncertainties from additional resonances, listed in Table \ref{additionalRes}, are determined by adding them to the best solution of PWA fit one-by-one. The differences between the fit yields determined with
and without the additional resonance are taken as systematic
errors. For the non-resonant mode $\psip\to\kk\pi^0$,
the uncertainty due to the $P$-wave $\kk$ system in the PWA fit is evaluated by replacing it with a $P$-wave $K\pi$
system. The difference in the fit yields is taken as a systematic error.

\item non-$\kk\pi^0$ background\\
The number of non-$\kk\pi$ background events is obtained from a $\pi^0$-sideband analysis and an exclusive MC
simulation. The difference in the signal yields corresponding to one standard deviation of this background is taken as a
systematic error.

\item the QED background\\
The QED background used at $\sqrt s=3.686$ GeV is produced via a MC simulation with amplitude information
obtained from a PWA fit to the data taken at $\sqrt s=3.773$ GeV. The uncertainty is estimated by replacing this
QED background with the continuum data taken at $\sqrt s=3.65$ GeV. The difference of the fitted yields between these
two approaches are 0.8\% and
9.9\% for $K^{*}_{2}$(1430) and $K^{*}$(892), respectively, and used as systematic uncertainties.

\item uncertainty of $K^*(1680)$ and $\rho(1700)$ widths\\
The decay widths of $K^{*}$(1680) and $\rho(1700)$ have large uncertainties; the world average values are
$\Gamma_{K^*(1680)}=322\pm110$ MeV and $\Gamma_{\rho(1700)}=250\pm100$ MeV \cite{pdg}.
The signal yields were re-obtained using widths that are changed by one standard deviation with respect to the nominal
value. The differences in signal yields between these two methods are taken as systematic errors.

\item uncertainties of masses and widths for the $K^*(892)$ and $K^*(1430)$\\
In the PWA fit, the masses and widths for the $K^*(892)$ and $K^*(1430)$ are fixed to the world average values. The differences in fit yields obtained by changing these parameters by one standard deviation are taken as systematic errors.

\end{enumerate}

All systematic errors from the PWA fit are listed in Table \ref{Sys:extra}.

Combining the systematic uncertainties from the PWA fit and the $\pi^0\kk$ event selection gives total systematic errors of $^{+8.3}_{-9.8}\%$ and $^{+15.6}_{-8.1}\%$ for $\psip\to K^{*}(892)^+ K^- +c.c.$ and $K^{*}_2(1430)^+K^{-}+c.c.$, respectively.
\begin{table}[htbp]
 \centering
 \caption{Summary of all systematic errors (\%).}
 \label{sys:gen_sys}
 \begin{tabular}{l|cccccc}
 \hline\hline
 Items               &$\pi^0\kk$    &$K^{*\pm}K^\mp$ &$K^{*\pm}_2 K^\mp$ &$\eta\kk$ &$\eta\phi$ &$\pi^{0}\phi$\\\hline
 Photon efficiency    &2              &2     &2      &2   &2    &2\\
 $\pi^0$ mass cut    &1.1            &1.1   &1.1    &--  &--   &--\\
 Kaon tracking       &2              &2     &2      &2   &2    &2\\
 PID                 &2              &2     &2      &2   &2    &2\\
 Kinematic fitting   &1.9            &3.2   &4.3    &2.1 &1.7  &2.1\\
 Number of ${\psip}$ decays    &4        &4  &4   &4&4 &4\\
 Background shape    &--             &--    &--     &1.6 &0.4  &--\\
 Fitting range       &--             &--    &--     &3.6 &0.6  &-- \\
 $Br[K^{*}_{J}]\to\pi^{0}K$    &--   &--    &2.4 &- &-  &-  \\
 $Br[P\to\gamma\gamma]$  &--   &--   &--    &0.5 &0.5  &---\\
 $Br[\phi\to KK]$    &-- &-- &-- &--  &1.2  &1.2\\
 QED background      &-- &-- &-- &--  &4.5  &--\\
 Additional states   &-- &-- &-- &4 &-- &--\\
 \hline
 Total               &6.3   &6.9 &6.2 &8.0 &7.3  &5.8 \\\hline\hline
 \end{tabular}
\end{table}

\begin{table}[htbp]
 \centering
 \caption{Summary of systematic uncertainties from the PWA (\%).}
  \label{Sys:extra}
 \begin{tabular}{l|ll}\hline\hline
 Sources                    &$K^{*}(892)^\pm K^\mp$        &$K^{*}_{2}(1430)^\pm K^\mp$ \\\hline
 Breit-Wigner                            &-0.1   &+0.9 \\
 Additional states          &$^{+5.2}_{-6.9}$ &$^{+10.3}_{-4.6}$ \\
 Non-$\kk\pi$ background    &$^{+1.4}_{-1.6}$   &$^{+1.2}_{-1.0}$\\
 QED background                          &-0.8   &+9.9  \\
 $K^*(1680),~\rho(1700)$ width &$^{+0.5}_{-1.1}$   &$^0_{-2.0}$\\
  $K^*(892),~K^*_2(1430)$
 Mass and width             &$^{+0.4}_{-0.4}$ &$^{+0.3}_0$\\
 Total                      &$^{+5.4}_{-7.3}$    &$^{+14.3}_{-5.1}$\\
 \hline\hline
 \end{tabular}
\end{table}
\section{Summary and discussion}
Using $(106\pm4)\times 10^6$ $\psip$ decays accumulated with BESIII,
we measured branching fractions for the $\psip\to K^{*}(892)^+
K^{-}+c.c,~K^{*}(1430)^+ K^{-}+c.c,~\eta\phi,~\pi^0\phi, \pi^0\kk$,
and $\eta \kk$ decays. The helicity forbidden decay $\psip\to
K^{*}_2(1430)^+K^{-}+c.c.$ is observed for the first time, and its
branching fraction is measured; this reflects a violation of the
helicity selection rule \cite{zhoaqiang}.  Table \ref{summary:table}
gives an overview of our results with comparisons with BESII- and
CLEO-measurements and world average values. The precision of our
measurements is better for all the modes, including a tightened upper
limit for $\pi^0\phi$. In the measurement of $Br(\psip\to\pi^0\kk)$,
all intermediate states are included in the branching fraction, while
for the measurement of $Br(\psip\to \kk\eta$),
$\psip\to\eta\phi$ is excluded. The measurements of branching
fractions for the $\psip\to K^{*}(892)^+K^{-}+c.c.$ and $\eta\phi$ are
consistent with BESII results within $1\sigma$, and CLEO measurements
within $2\sigma$.

Using the world average values of branching fractions for $\jp$
decays, the $Q_h$ values are calculated and listed in Table
\ref{summary:table}. For $\psip\to K^{*}(892)^+K^-+c.c$ and
$\eta\phi$, the $Q_h$ values significantly deviate from the expected
value of 12\%.

\begin{table}[htbp]
  \centering\footnotesize
  \caption{Summary of the measured branching fractions compared with PDG \cite{pdg} values, together with CLEO \cite{cleoVP} and BESII \cite{bes2VP} measurements. The upper limit is given at the 90\% confidence level. The first error is statistical, and the second error is systematic. Here $\epsilon,~\textrm{N}^{obs}$, and Br denote the detection efficiency, the number of observed events, and the branching fraction, respectively. The variable $Q_h$ is defined by Eq. (\ref{qh_definition}).}
  \label{summary:table}
  \begin{tabular}{l|llll|lll}\hline\hline
Mode($\psip\to$)             &$\epsilon$(\%)  &N$^{obs}$   &Br$(\times10^{-5})$       &$Q_h$ (\%) & PDG($\times10^{-5}$) & CLEO($\times10^{-5}$) & BESII($\times10^{-5}$) \\\hline
  $\pi^0 \kk$(inclusive)      &21.52  &917$\pm$37          &4.07$\pm0.16\pm0.26$              &--- & $<8.9$ \cite{kkpi0} & --- & ---\\
  $ K^{*}(892)^+K^-+c.c.$     &20.25  &224$\pm$21 &3.18$\pm0.30^{+0.26}_{-0.31}$            &$0.62\pm0.09$ & $1.7^{+0.8}_{-0.9}$ & $1.3\pm1.0\pm0.3$ & $2.9\pm1.3\pm0.4$\\
  $K^{*}_2(1430)^+ K^-+c.c.$   &20.28  &251$\pm$22 &7.12$\pm0.62^{+1.13}_{-0.61}$            &$>2$ & --- & --- &---\\
  $\eta \kk$($\eta\phi$ excluded)                  &22.10  &284$\pm$27       &3.08$\pm0.29\pm0.25$             &--- & $<13$ & $<13$ & ---\\
  $\eta\phi$                 &33.53  &216$\pm$16         &3.14$\pm0.23\pm0.23$             &$4.19\pm0.61$ & $2.8^{+1.0}_{-0.8}$ & $2.0\pm1.1\pm0.4$ & $3.3\pm1.1\pm0.5$\\
  $\pi^{0}\phi$              &35.63  &$<$10                &$<$0.04                         &---& $<0.4$ & $0.7$ &$0.4$\\\hline\hline
  \end{tabular}
\end{table}

\acknowledgements
The BESIII collaboration thanks the staff of BEPCII and the computing center for their hard efforts. This work is supported in part by the Ministry of Science and Technology of China under Contract No. 2009CB825200; National Natural Science Foundation of China (NSFC) under Contracts Nos. 10625524, 10821063, 10825524, 10835001, 10935007; Joint Funds of the National Natural Science Foundation of China under Contract No. 11079008; the Chinese Academy of Sciences (CAS) Large-Scale Scientific Facility Program; CAS under Contracts Nos. KJCX2-YW-N29, KJCX2-YW-N45; 100 Talents Program of CAS; Istituto Nazionale di Fisica Nucleare, Italy; Siberian Branch of Russian Academy of Science, joint project No 32 with CAS; U. S. Department of Energy under Contracts Nos. DE-FG02-04ER41291, DE-FG02-91ER40682, DE-FG02-94ER40823; U.S. National Science Foundation; University of Groningen (RuG) and the Helmholtzzentrum fuer Schwerionenforschung GmbH (GSI), Darmstadt; WCU Program of National Research Foundation of Korea under Contract No. R32-2008-000-10155-0.
This paper is also supported by the NSFC under Contract Nos. 10979038,
  10875113, 10847001, 11005115; Innovation Project of Youth Foundation of Institute of
  High Energy Physics under Contract No. H95461B0U2.

\end{document}